\documentclass[twocolumn,preprintnumbers,amsmath,amssymb,prl]{revtex4}
\usepackage{graphicx}% Include figure files
\usepackage{dcolumn}% Align table columns on decimal point
\usepackage{bm}% bold math
\usepackage{color}
\usepackage{ulem}
\begin{document}

\title{
Evidence for an FFLO state with segmented vortices in the BCS-BEC-crossover superconductor FeSe
}

\author{S.\,Kasahara$^{1,*}$}
\author{Y.\,Sato$^{1,*}$}
\author{S.\,Licciardello$^2$}
\author{M.\,\v{C}ulo$^2$}
\author{S.\,Arsenijevi\'{c}$^3$}
\author{T.\,Ottenbros$^2$}
\author{T.\,Tominaga$^1$}
\author{J.\,B\"{o}ker$^4$}
\author{I.\,Eremin$^4$}
\author{T.\,Shibauchi$^5$}
\author{J.\,Wosnitza$^{3,6}$}
\author{N.\,E. Hussey$^2$}
\author{Y.\,Matsuda$^1$}

\affiliation{$^1$Department of Physics, Kyoto University, Kyoto 606-8502 Japan}
\affiliation{$^2$High Field Magnet Laboratory (HFML-EMFL) and Institute for Molecules and Materials, Radboud University, Nijmegen, The Netherlands}
\affiliation{$^3$Hochfeld-Magnetlabor Dresden (HLD-EMFL) and W\"urzburg-Dresden
Cluster of Excellence ct.qmat, Helmholtz-Zentrum Dresden-Rossendorf, D-01328 Dresden, Germany}
\affiliation{$^4$Institut f\"{u}r Theoretische Physik III, Ruhr-Universit\"{a}t Bochum, D-44801 Bochum, Germany}
\affiliation{$^5$Department of Advanced Materials Science, University of Tokyo, Chiba 277-8561, Japan}
\affiliation{$^6$Institut f\"ur Festk\"orper- und Materialphysik, Technische Universit\"at Dresden, 01062 Dresden, Germany}

%\date{\today}%

%\pacs{}


\begin{abstract}
We present resistivity and thermal-conductivity measurements of superconducting FeSe in intense magnetic fields up to 35\,T applied parallel to the $ab$ plane. At low temperatures, the upper critical field $\mu_0 H_{c2}^{ab}$ shows an anomalous upturn, while thermal conductivity exhibits a discontinuous jump at $\mu_0 H^{\ast}\approx 24$\,T well below $\mu_0 H_{c2}^{ab}$, indicating a first-order phase transition in the superconducting state. This demonstrates the emergence of a distinct field-induced superconducting phase. Moreover, the broad resistive transition at high temperatures abruptly becomes sharp upon entering the high-field phase, indicating a dramatic change of the magnetic-flux properties. We attribute the high-field phase to the Fulde-Ferrel-Larkin-Ovchinnikov (FFLO) state, where the formation of planar nodes gives rise to a segmentation of the flux-line lattice. We point out that strongly orbital-dependent pairing as well as spin-orbit interactions, the multiband nature, and the extremely small Fermi energy are important for the formation of the FFLO state in FeSe.
\end{abstract}

\maketitle

Exotic superconductivity with a nontrivial Cooper-pairing state has been a longstanding issue of interest in condensed-matter physics.  Among possible exotic states, a spatially nonuniform superconducting state in the presence of strong magnetic fields caused by the paramagnetism of conduction electrons has been the subject of great interest  after the pioneering work by Fulde and Ferrell as well as Larkin and Ovchinnikov (FFLO) \cite{FFLO, LO}. In the FFLO state, pair breaking due to the Pauli paramagnetic effect is reduced by forming a new pairing state ({\boldmath $k$}$\uparrow$, {\boldmath $-k+q$}$\downarrow$) with {\boldmath $|q|$} $\sim g\mu_B H/\hbar \upsilon_F$ ($\upsilon_F$ is the Fermi velocity, $g$ the $g$-factor, and $\mu_B$ the Bohr magneton) between Zeeman split parts of the Fermi surface, instead of ({\boldmath $k$}$\uparrow$, {\boldmath $-k$}$\downarrow$) pairing in BCS superconductors [Figs.\,1(a) and 1(b)]. The fascinating aspect of the FFLO state is that the superconducting order parameter, in its simplest form, is modulated as $\Delta \propto \sin$ {\boldmath $q\cdot r$}, and periodic planar nodes appear perpendicular to the magnetic field near the upper critical field $H_{c2}$, leading to a segmentation of the vortices into pieces of length $\Lambda=\pi/${\boldmath $|q|$} [Fig.\,1(c)].

Despite tremendous efforts in the search for the FFLO states in the past half century, indications of its experimental realization have been reported in only a few candidate materials, including quasi-two-dimensional (2D) organic superconductors 
and the heavy-fermion superconductor CeCoIn$_5$ \cite{ZwicknaglBCS50, Wosnitza18,  Shimahara}. In both systems, a thermodynamic phase transition occurs below $H_{c2}$ and a high-field superconducting phase emerges at low temperatures \cite{Radvan03,Bianchi03,Lortz07,Agosta17}. In the former, each superconducting layer is very weakly coupled via the Josephson effect.  The FFLO state is observed in a magnetic field $\bm{H}$ applied parallel to the layers, where the magnetic flux is concentrated in the regions between the layers forming coreless Josephson vortices. Therefore, the segmentation of the vortices by FFLO nodes, which is one of the most fascinating properties of the FFLO state, is not expected.
The presence of the FFLO phase in CeCoIn$_5$, on the other hand, remains a controversial issue.
Magnetic order occurs in the high-field phase \cite{Kenzelmann08}, indicating that this phase is not a simple FFLO phase. Although the coexistence of FFLO and spin- or pair-density-wave states has been proposed \cite{Agterberg08, Agterberg09, Yanase09, Hatakeyama15}, the nature of the superconducting order parameter remains open. 
\textcolor{black}{Possible FFLO states have also been discussed in CeCu$_2$Si$_2$ and KFe$_2$As$_2$~\cite{Kitagawa18,Cho17}. In the former, however, no phase transition line has reported in the superconducting phase. In the latter, the high-field phase disappears when the magnetic field is very slightly tilted away from the $ab$ plane. 
It is not clear whether such a behavior is compatible with the FFLO state in a superconductor whose anisotropy is much smaller than Josephson coupled 2D organic compounds.
}
For a deeper understanding of the FFLO pairing state, further superconductors revealing this state are strongly required.

The layered iron-chalcogenide superconductor FeSe ($T_c\approx9$\,K) has aroused enormous enthusiasm to study the
exotic superconductivity with various distinct features \cite{Hsu08,Coldea17,Boehmer18}. FeSe is a compensated semimetal, which exhibits a structural transition from tetragonal to orthorhombic crystal symmetry at $T_s\approx 90$\,K \cite{McQueen09}. %, 
In contrast to other iron-based compounds, no magnetic order occurs below $T_s$ \cite{Imai09, Baek15, Boehmer15}.  The Fermi surface of FeSe consists of hole cylinders around the zone center and compensating electron cylinders around the zone corner [Fig.\,1(d)] \cite{Coldea17, Terashima14, Suzuki15, Watson15, Onari16, Yi19}. The superconducting gap function is highly anisotropic \cite{Song11, Kasahara14, Bourg16, Watashige15, Watashige17, Sato18, Hanaguri18}. Recent angle-resolved photoemission-spectroscopy and quasi-particle-interference experiments reported that the gap function has nodes or deep minima at the long axis of the elliptical hole pocket [Fig.\,1(e)] \cite{Xu16, Hashimoto18,Sprau17,Liu18}.%, 

\begin{figure}[t]
\begin{center}
\includegraphics[width=1.0\linewidth]{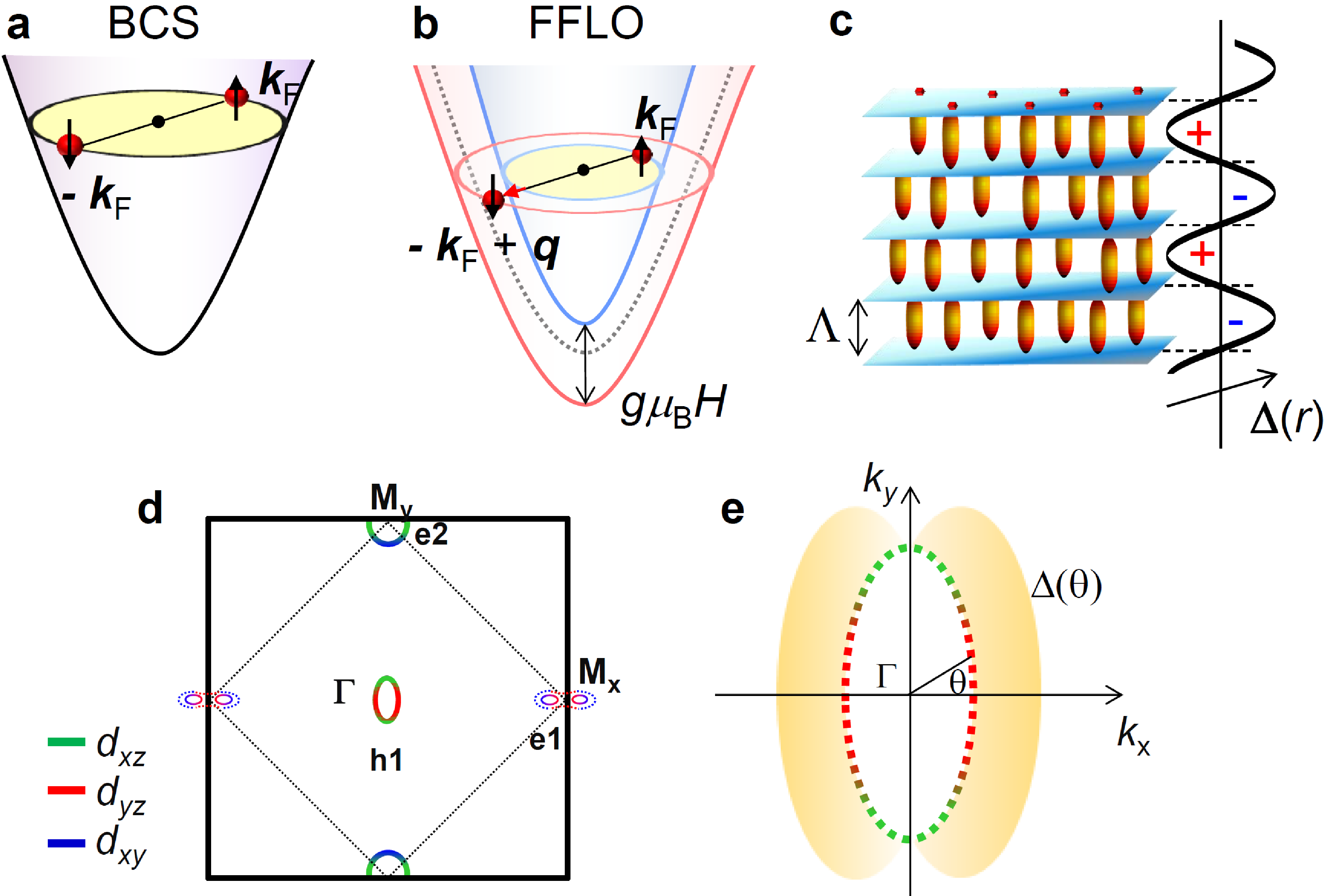}
\caption{(a) Schematic illustration of Cooper pairing ({\boldmath $k$}$\uparrow$, {\boldmath $-k$}$\downarrow$) in the BCS state.  (b) Pairing state with ({\boldmath $k$}$\uparrow$, {\boldmath $-k+q$}$\downarrow$) in the FFLO state. (c) Schematic illustration of the superconducting order parameter $\Delta$ in real space and segmentation of the magnetic flux lines by planar nodes. (d) Schematic figure of the Fermi surface of FeSe in the nematic state. Green, red, and blue areas represent the Fermi-surface regions dominated by $d_{xz}$, $d_{yz}$, and $d_{xy}$ orbitals, respectively. (e) Angular dependence of $\Delta$ at the hole pocket of FeSe, where $\theta$ is the angle from the $k_x$ axis. The superconducting gap is highly orbital dependent and nodes or deep minima appear at $\theta = \pm 90^\circ$. }
 \end{center}
 \end{figure}

In FeSe, the presence of a high-field phase has been suggested by a kink anomaly of the thermal conductivity, $\kappa$, below $H_{c2}$ in perpendicular field ({\boldmath $H$} $\parallel c$) \cite{Kasahara14}. Although this high-field phase has been discussed in terms of a possible FFLO state \cite{Kasahara14,Watashige17,Song18,Song19}, it is an open question what kind of state is actually realized. Therefore, it is important to investigate the superconducting state in parallel field ({\boldmath $H$}$\parallel ab$), in which the superconductivity survives up to a higher field.  In this Letter, we report measurements of the in-plane electrical resistivity, $\rho$, and $\kappa$ of FeSe in parallel field up to 35\,T.  We provide compelling evidence of a distinct high-field superconducting phase, which is separated %by the first order phase transition
from the low-field phase via a first-order phase transition. We attribute this high-field phase to an FFLO state, in which the Abrikosov flux-line lattice is segmented by periodic nodal planes. We point out that the peculiar electronic structure of FeSe is primarily responsible for the FFLO formation.

High-quality single crystals of FeSe are grown by chemical vapor-transport technique \cite{Bohmer13}. %
Measurements of $\kappa$ are conducted at the High Field Magnetic Laboratory in Nijmegen using a specially built sample holder \cite{Arsenijevic16}. 
Since our crystal is twinned, {\boldmath $H$} is applied along the diagonal direction in the $ab$ plane ({\boldmath $H$} $\parallel [110]_{\rm O}$, in orthorhombic notation), so that
two orthorhombic domains yield the same response to {\boldmath $H$}.  Electrical and thermal currents, {\boldmath $j$} and {\boldmath $j$}$_{Th}$, respectively, are applied parallel to {\boldmath $H$}.

\begin{figure}[t]
\begin{center}
\includegraphics[width=1.0\linewidth]{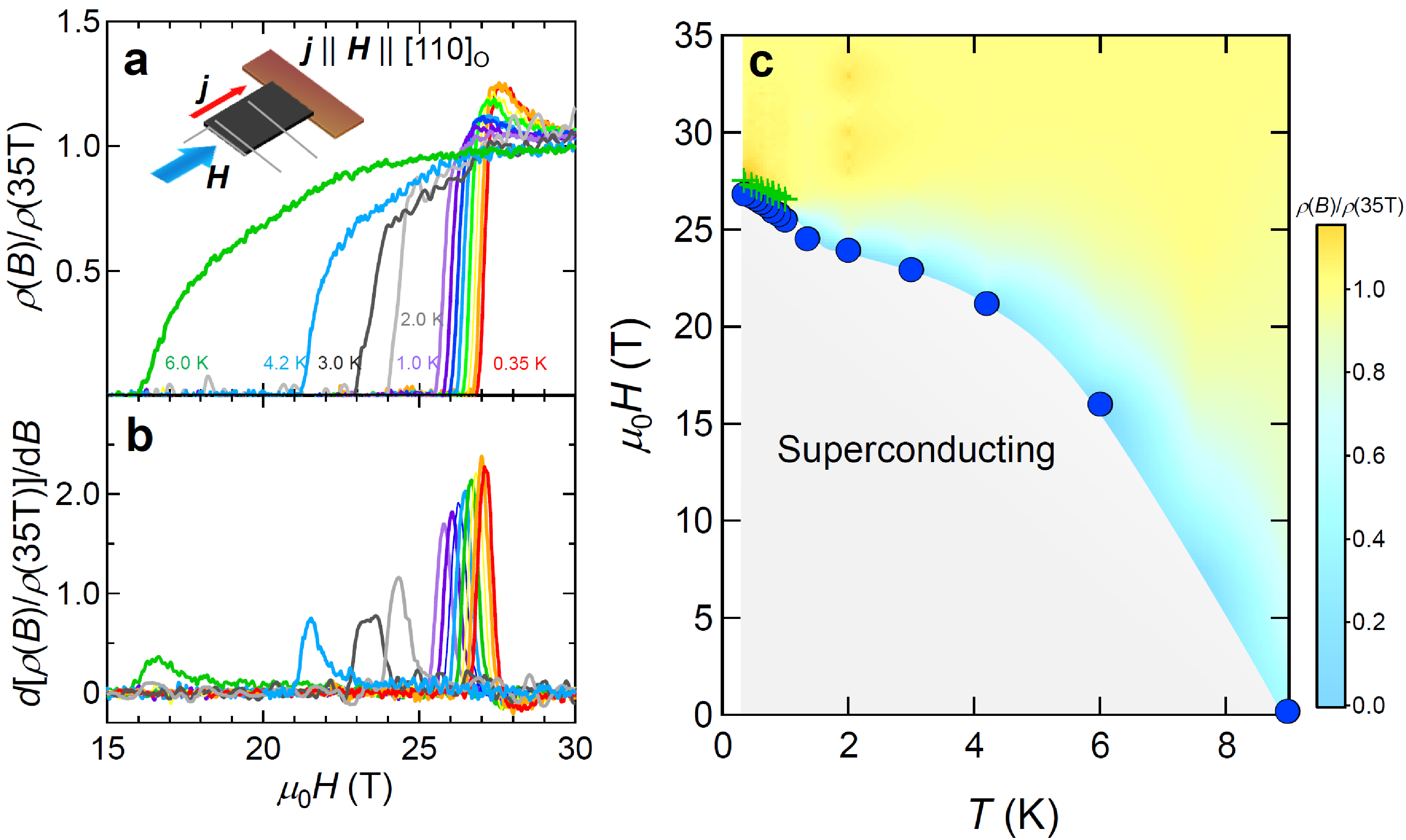}
\caption{(a) Magnetic-field dependence of the in-plane resistivity normalized by the value at $\mu_0H$=35\,T, $\rho(H)/\rho$(35\,T), and (b) its field derivative up to 30\,T, respectively. (See also Supplemental Material.) The broad transition at high temperature abruptly becomes sharp at low temperatures. (c) Field-temperature ($H$-$T$) phase diagram of FeSe for {\boldmath $H$} applied in the $ab$ plane. The blue circles show the irreversibility field, $H_{irr}$ where finite resistance first appears. The color plot represents the magnitude of $\rho(H)/\rho$(35\,T) above the superconducting transition.  The green crosses represent the field $H_{p}$ at which $\rho(H)/\rho(35\,{\rm T})$ shows a maximum. 
}
\end{center}
\end{figure}

Figures 2(a) and 2(b) depict the field dependence of the resistivity normalized by the value at $\mu_0H$ = 35\,T, $\rho(H)/\rho$(35\,T), and its field derivative, respectively (see also Supplemental Material and Ref.~\cite{Licciardello19}.) There are several remarkable features.
The resistive transition in magnetic field, which exhibits a significant broadening at high temperatures,  becomes sharp below $\sim 1$\,K. The broad resistive transition suggests a strongly fluctuating superconducting order parameter~\cite{Kasahara16}, which gives rise to the drift motion of vortices in the liquid state.
The onset field of non-zero resistivity is the irreversibility field, $H_{irr}$, that marks the vortex solid-liquid transition.

Figure 2(c) depicts the $T$ dependence of $H_{irr}$ (filled blue circles) along with a color plot illustrating the magnitude of $\rho(H)/\rho(35\,{\rm T})$. 
Above $T \sim 1$\,K, the in-plane upper critical field $H_{c2}^{ab}$ is expected to be located well above $H_{irr}$, although no feature is observed in the measured resistivity. On the other hand, below $\sim1$\,K, where the sharp resistive transition is observed, $H_{c2}^{ab}$ is expected to be close to $H_{irr}$.  Therefore, we can safely conclude that $H_{c2}^{ab}$ exhibits an anomalous upturn below $\sim2$\,K, suggesting the formation of a  high-field superconducting phase.

\begin{figure}[t]
\begin{center}
\includegraphics[width=1.0\linewidth]{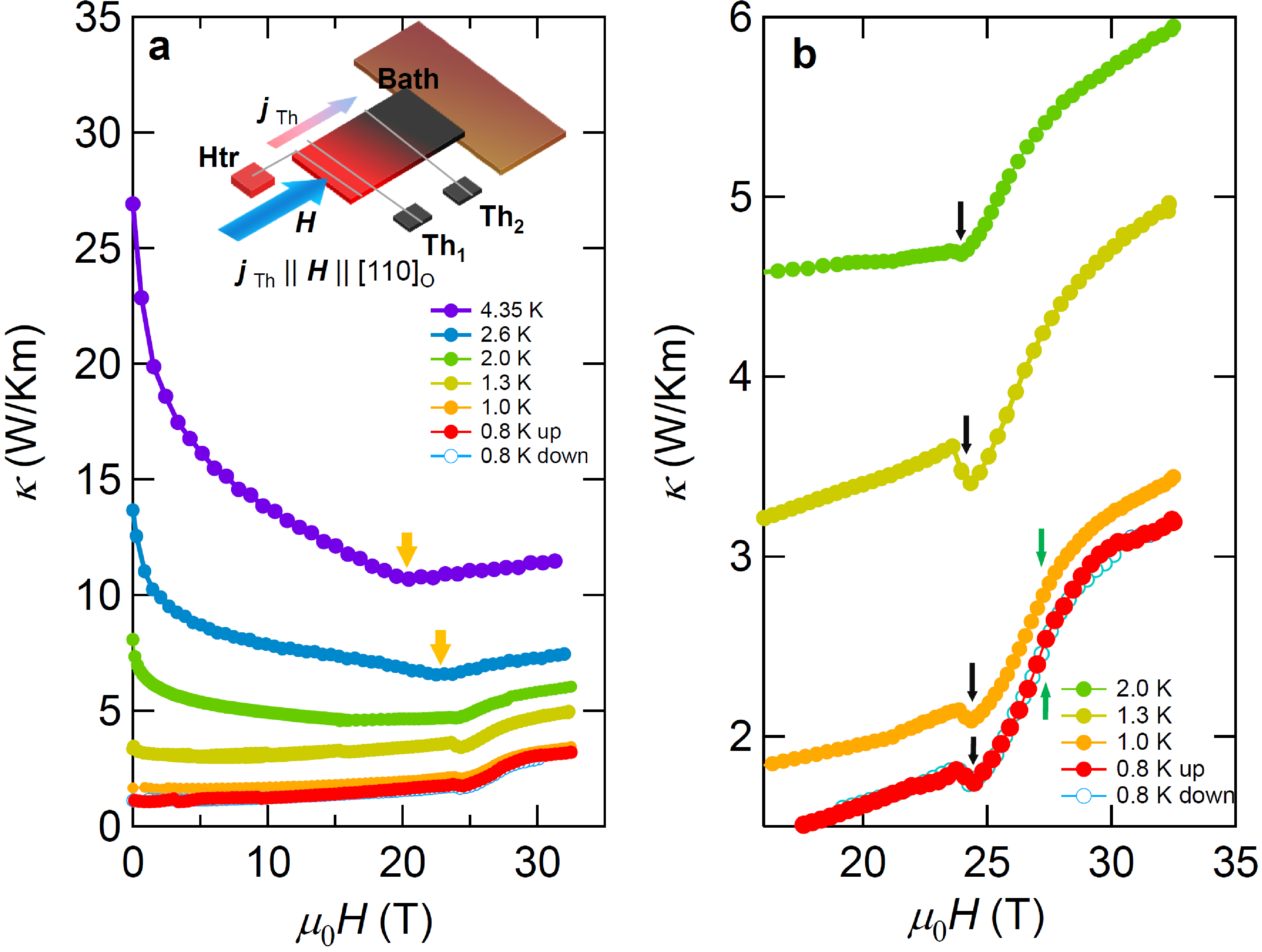}
\caption{(a) Magnetic-field dependence of the thermal conductivity in FeSe for $\bm{H}\parallel ab$. The inset shows a schematic illustration of the experimental set-up of the thermal-conductivity measurements. Orange arrows indicate the magnetic field $H_k$ at which a kink-like minimum of $\kappa(H)$ appears. (b) The same data below $T = 2$\,K plotted for the high-field regime above $\mu_0H = 16$\,T. A discontinuous downward jump at $\mu_0H = 24$\,T appears inside the superconducting state as indicated by the black arrows. Green arrows indicate the field $H_{p}$ determined by our resistivity measurements. }
 \end{center}
 \end{figure}

The presence of an anomalous high-field phase is confirmed by thermal-conductivity measurements. Figure 3(a) shows the $H$ dependence of $\kappa$ up to 33\,T. Above $\sim2$\,K, $\kappa(H)$ first decreases with $H$ and then increases gradually after attaining a kink-like minimum at $\mu_0H_{k}=20$\,T and 20.5\,T at 4.35\,K and 2.6\,K, respectively, which are close to $H_{irr}$. The initial reduction of $\kappa(H)$ is caused by the suppression of the quasiparticle mean free path due to  introduction of vortices \cite{Krishana97,Franz99,Izawa01,Kasahara05, Kasahara14}. Below $T \sim 1$\,K, $\kappa(H)$  increases with $H$ without showing an initial reduction.
Figure 3(b) displays $\kappa(H)$ below 2.0\,K and above 16\,T.

The most remarkable feature of the low-$T$ data is that $\kappa(H)$ exhibits a discontinuous downward jump at $\mu_0H^{\ast}\approx 24$\,T (black arrows).  At $H^{\ast}$, $\kappa(H)$ shows a large change of the field slope and increases steeply with $H$ above $H^{\ast}$. It should be stressed that $H^{\ast}$ is deep inside the superconducting state at low temperature, as evidenced by the fact that $H^{\ast}$ is well below $H_{irr}$.  Figure 4 displays the $T$ dependence of $H_{irr}$ and $H^{\ast}$.  As the temperature is increased, $H^{\ast}$ decreases gradually and coincides with $H_{irr}$ at about 2\,K. Note that the jump of $\kappa(H)$, which is intimately related to a jump in entropy, is a strong indication of a first-order phase transition, as reported for CeCoIn$_5$ and URu$_2$Si$_2$ \cite{Izawa01, Kasahara05, Kasahara07, Kim16}.  No discernible anomaly of $\kappa(H)$ is observed above about 2\,K, indicating that the first-order transition occurs only within the superconducting state. Thus, our $\kappa(H)$ measurements provide strong evidence for a distinct high-field superconducting phase, which is separated by a first-order phase transition from the superconducting low-field phase.

\begin{figure}[t]
\begin{center}
\includegraphics[width=0.8\linewidth]{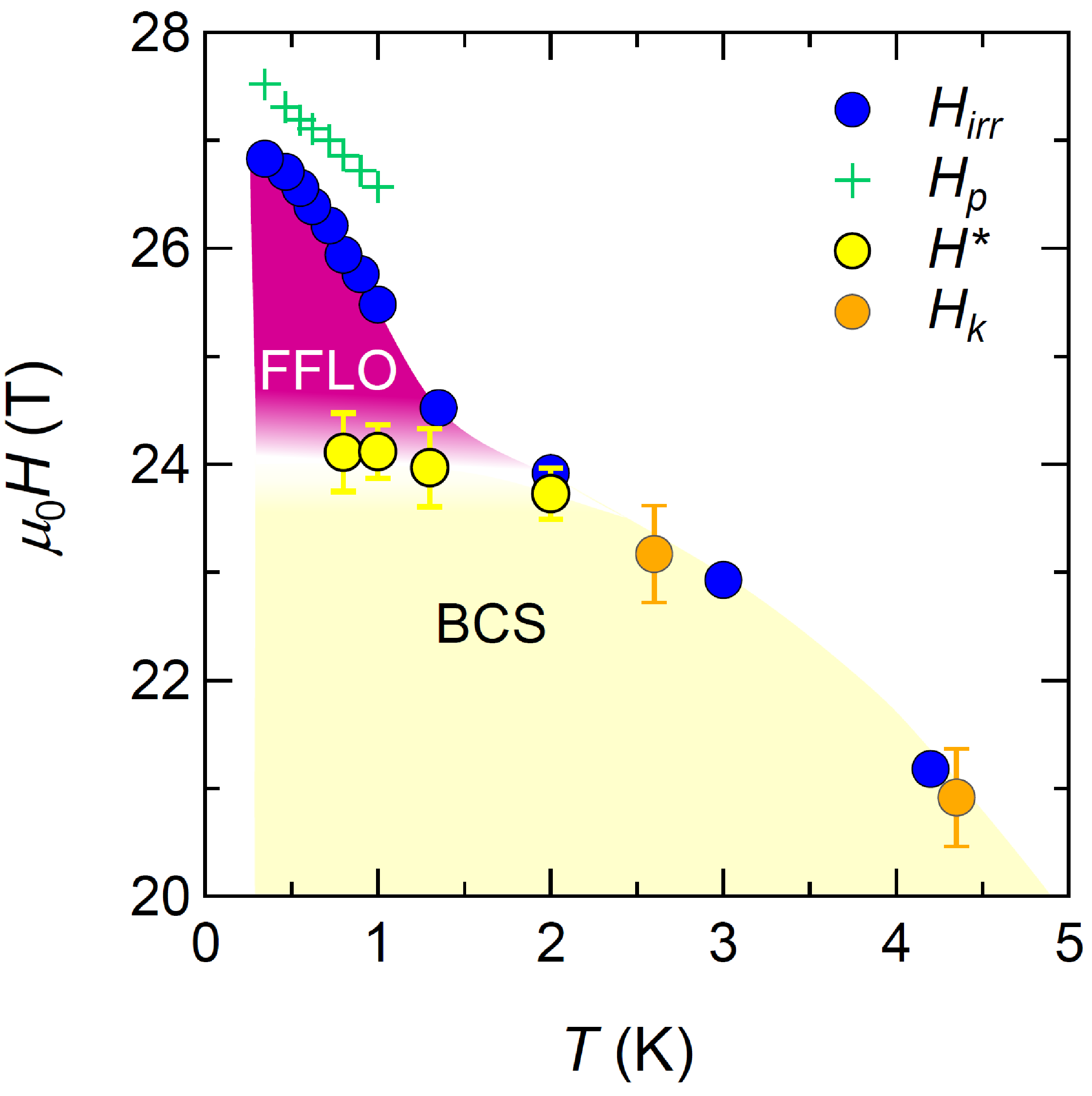}
\caption{High-field phase diagram of FeSe for $\bm{H}\parallel ab$ plane. Blue circles and green crosses show $H_{irr}$ and  $H_{p}$ determined by resistivity measurements. Orange and yellow circles show $H_k$ and $H^{\ast}$ determined by thermal-conductivity measurements, respectively. Above the first-order phase transition field $H^{\ast}$, a distinct field-induced superconducting phase emerges at low temperatures.}
\end{center}
\end{figure}

We  point out that the high-field superconducting phase  is not an antiferromagnetic (AFM) ordered phase.  When such order occurs, the Fermi surface is folded into the (reduced) AFM Brillouin zone, and, as a result, a partial energy gap opens over portions of the Fermi surface. However, quantum-oscillation measurements show no evidence of such a band folding \cite{Terashima14}.  Moreover, given its %in FeSe with
very small Fermi surfaces, $H_{c2}$ in FeSe is expected to be largely suppressed by AFM ordering due to a concomitant reduction in the charge-carrier number.

We associate the high-field phase with an FFLO phase for the following reasons. Firstly, the $H$-$T$ phase diagram shown in Fig.\ 4, including the steep enhancement of $H_{c2}^{ab}$ at low temperature and the first-order phase transition at a largely $T$ independent $H^{\ast}$ 
bears a striking resemblance to that expected for the FFLO transition \cite{Shimahara,Wosnitza18}.
Secondly, the FFLO state requires a large Maki parameter (ratio of the orbital to the Pauli-paramagnetic limiting fields)
$\alpha_M\equiv \sqrt{2}H_{c2}^{orb}/H_{c2}^P>1.5$~\cite{Saint-James}.
In FeSe, the Fermi energies of both hole and electron pockets are extremely small, leading to large ratios of the superconducting energy gap to the Fermi energy, $\Delta_h/\varepsilon_F \approx 0.3$ for the hole band and $\Delta_e/\varepsilon_F^e \approx 0.5$ for the electron band \cite{Kasahara14,Hanaguri18}. This places FeSe deep inside the so-called BCS-BEC crossover regime, where the extent of the Cooper pairs is comparable to the average distance between electrons \cite{Kasahara14,Kasahara16,Hanaguri19,LeeJC}. Using $\alpha_M\equiv \sqrt{2}H_{c2}^{orb}/H_{c2}^P\approx 2m^{\ast}/m_e\cdot \Delta/\varepsilon_F$ in the BCS limit, where $m^{\ast}$ and $m_e$ are the effective and free electron mass, respectively, $\alpha_M$ is found to be as large as $\sim 5$ and $\sim 2.5$ for the hole and electron pockets, respectively. 
\textcolor{black}{In addition, the present crystal of FeSe is in the ultra-clean limit with extraordinary long mean free path $\ell$ (See Supplemental Material). Such large values of $\alpha_M$ and $\ell$ are the prerequisites for the realization of the FFLO state.}
\textcolor{black}{Thirdly, planar nodes perpendicular to $\bm{H}$ are expected as the most optimal solution for the lowest Landau level. In the present geometry, where {\boldmath $j$}$_T$ $\parallel$ {\boldmath $H$}, quasiparticles that conduct heat are expected to be scattered by the periodic planar nodes upon entering the FFLO phase.}
This leads to a reduction of $\kappa(H)$ just above $H^*$, which is consistent with the present results.
Finally, as the $c$-axis coherence length ($\xi_c \approx 1.3$\,nm) well exceeds the interlayer distance (0.55\,nm) ~\cite{Hsu08,Terashima14}, one-dimensional tube-like Abrikosov vortices are formed even in a parallel field. In this case, the planar node formation leads to a segmentation of the vortices into pieces of length $\Lambda$. The pieces are largely decoupled and, hence, better able than conventional vortices to position themselves at pinning centers, %giving rise to the
leading to an enhancement of the pinning forces of the flux lines in the FFLO phase.  This is consistent with the observed sharp resistive transition above $H^{\ast}$.

One intriguing feature of the high-field phase is that $\rho(H)$ exhibits an anomalous enhancement from the normal-state value just above $H_{c2}^{ab}$ [Fig.\ 2(a)]. The origin of this enhancement is not clear. As shown by green crosses in Fig.\ 2(c), which indicate the field at which $\rho(H)/\rho(35\,{\rm T})$ shows a maximum, this enhancement occurs slightly above the high-field phase. Therefore, it is tempting to consider that the enhancement is related to a peculiar electronic state above the FFLO transition.  Its clarification deserves further investigations.

Theoretically, the multi-orbital nature \cite{Gurevich10, Takahashi14, Adachi15}, nematicity, small Fermi energies (in comparison with the superconducting pairing scale)
and an effectively strong spin-orbit coupling, $\lambda_{so} \sim \varepsilon_F$, make the analysis of the FFLO state for {\boldmath $H$}$\parallel ab$ interesting and challenging.  
We point out that large spin-orbit coupling plays an important role for the FFLO formation in FeSe by inspection of the effective $g$-factors for {\boldmath $H$}$\parallel ab$ and {\boldmath $H$}$\parallel c$ 
for the hole and electron pockets. Here, we adopt the  band structure of FeSe obtained by the orbitally projected model \cite{Kang18} and include spin-orbit coupling as well as the nematic order (see Supplemental Material for details). Due to spin-orbit coupling, the Zeeman field acts differently on the hole and electron pockets, and is asymmetric for {\boldmath $H$}$\parallel ab$ and {\boldmath $H$}$\parallel c$.  In particular, for the hole pocket, spin-orbit coupling acts as an imaginary pseudo-Zeeman field along the $z$ direction. As a result, the true Zeeman field along $z$ further splits the remaining pocket by an amount $\pm g \mu_B H_z$, while along $x$, 
it acquires an effective reduced $g$-factor, $g_x^{\Gamma} < g$.  For the electron pockets, the situation is even more complex due to the involvement of the $xy$ and $yz$ orbitals, and the corresponding $g$-factors are reduced for both orientations of the Zeeman field due to spin-orbit coupling. For {\boldmath $H$}$\parallel c$, $g_{z}^{M}$ vanishes at the crossing points of two dispersions, yielding no Zeeman splitting there. By contrast, for {\boldmath $H$}$\parallel ab$, the splitting is reduced, yet the effective $g_x^{M}$ is finite everywhere (see Supplemental Material).

Note that, in the iron-based superconductors, it is believed that interband scattering of Cooper pairs of predominantly $yz$-orbital character from hole to electron pockets plays an important role.  For {\boldmath $H$}$\parallel c$, the simple analysis of the Zeeman field on the Fermi-surface pockets indicates that the splitting on the hole pocket is large while that on the electron pocket is much smaller, yielding a large momentum mismatch for scattering of the FFLO pairs.
In contrast, such a mismatch is much smaller for {\boldmath $H$}$\parallel ab$, as the effective $g$-factor is reduced in both pockets due to spin-orbit coupling. Therefore, the formation of the FFLO state is more favored for {\boldmath $H$}$\parallel ab$ than for {\boldmath $H$}$\parallel c$.  Moreover, the magnitude of the spin imbalance introduced through Zeeman splitting in magnetic field, $P=(N_{\uparrow}-N_{\downarrow})/N_{\uparrow}+N_{\downarrow}) \approx g\mu_BH/\varepsilon_F$.
Here, $N_\uparrow$ and $N_\downarrow$ are the numbers of up and down spins, respectively. In almost all superconductors, $P$ is very small, i.e., $P\sim 10^{-3}$--$10^{-2}$  even near $H_{c2}$.  In FeSe in the BCS-BEC crossover regime, the Zeeman effect is particularly effective in shrinking the Fermi volume associated with the spin minority, giving rise to a highly spin-imbalanced phase.  Near $H_{c2}$ for {\boldmath $H$}$\parallel c$, $\varepsilon_F^e\sim 4$\,meV yields $P\sim 0.4$ for electron pockets, assuming $g\sim2$,  indicating that electron pockets are highly polarized.  It is questionable that superconducting pairing is induced in such an extremely polarized state.  These considerations suggest that the high-field phase for {\boldmath $H$}$\parallel c$ may  not be an FFLO state.

It has been shown that the FFLO instability is  sensitive to the nesting properties of the Fermi surface. When the Fermi surfaces have flat parts, the FFLO state is more stabilized through nesting \cite{Wosnitza18}. As the portion of the hole pocket derived from the $d_{yz}$ orbital forms a Fermi-surface sheet that is more flattened than the other portion of the Fermi surface, this 1D-like Fermi sheet is likely to be responsible for the FFLO state [Fig.\ 1(d)].
The determination of the relevant {\boldmath $q$}-vector is crucially important for clarifying the orbital selective FFLO pairing. %

In summary, we demonstrate the presence of  a distinct low-temperature and high-field superconducting phase that is accessed through a first-order phase transition in parallel field. In this high-field phase, the upper critical field increases with a steep upward slope as the temperature is lowered and the magnetic-flux properties change dramatically. We attribute the high-field phase to an FFLO state. Furthermore, we speculate that the
strongly orbital-dependent pairing interaction and spin-orbit coupling, as well as the multiband BCS-BEC crossover nature are the essential ingredients
for the formation of an FFLO state in FeSe. The high-field  phase in FeSe provides the first genuine opportunity
to study a segmentation of the flux-line lattice by periodic nodal planes.

We thank M. Houzet, J.\,S. Kim, Y. Yanase for stimulating discussion. This work is supported by Grants-in-Aid for Scientific Research (KAKENHI) (Nos. 15H02106, 15H03688, 15KK0160, 18H01177, 18H05227, 19H00649) and on Innovative Areas "Topological Material Science" (No. 15H05852) "Quantum Liquid Crystals" (No. 19H05824) from the Japan Society for the Promotion of Science. We acknowledge support from the Deutsche Forschungsgemeinschaft (DFG) through the W\"urzburg-Dresden Cluster of Excellence on Complexity and Topology in Quantum Matter $ct.qmat$ (EXC 2147, project-id 39085490), the ANR-DFG grant Fermi-NESt, and by HFML-RU and HLD-HZDR, members of the European Magnetic Field Laboratory (EMFL).

\end{document}